\documentclass[onecolumn,amsmath,nofootinbib,showpacs]{revtex4-1}
\usepackage{graphicx}
\usepackage{dcolumn}
\usepackage{bm}
\usepackage{color}
\begin{document} 
\newcommand{\hs}{\hspace*{0.5cm}}
\newcommand{\vs}{\vspace*{0.5cm}}
\newcommand{\be}{\begin{equation}}
\newcommand{\ee}{\end{equation}}
\newcommand{\bea}{\begin{eqnarray}}
\newcommand{\eea}{\end{eqnarray}}
\newcommand{\ben}{\begin{enumerate}}
\newcommand{\een}{\end{enumerate}}
\newcommand{\bwt}{\begin{widetext}}
\newcommand{\ewt}{\end{widetext}}
\newcommand{\nn}{\nonumber}
\newcommand{\crn}{\nonumber \\}
\newcommand{\Tr}{\mathrm{Tr}}
\newcommand{\non}{\nonumber}
\newcommand{\noi}{\noindent}
\newcommand{\al}{\alpha}
\newcommand{\la}{\lambda}
\newcommand{\bet}{\beta}
\newcommand{\ga}{\gamma}
\newcommand{\va}{\varphi}
\newcommand{\om}{\omega}
\newcommand{\pa}{\partial}
\newcommand{\+}{\dagger}
\newcommand{\fr}{\frac}
\newcommand{\bc}{\begin{center}}
\newcommand{\ec}{\end{center}}
\newcommand{\Ga}{\Gamma}
\newcommand{\de}{\delta}
\newcommand{\De}{\Delta}
\newcommand{\ep}{\epsilon}
\newcommand{\varep}{\varepsilon}
\newcommand{\ka}{\kappa}
\newcommand{\La}{\Lambda}
\newcommand{\si}{\sigma}
\newcommand{\Si}{\Sigma}
\newcommand{\ta}{\tau}
\newcommand{\up}{\upsilon}
\newcommand{\Up}{\Upsilon}
\newcommand{\ze}{\zeta}
\newcommand{\ps}{\psi}
\newcommand{\Ps}{\Psi}
\newcommand{\ph}{\phi}
\newcommand{\vph}{\varphi}
\newcommand{\Ph}{\Phi}
\newcommand{\Om}{\Omega}
 
\title{\boldmath Asymmetric matter from $B-L$ symmetry breaking}

\author{Phung Van Dong}
\email{Corresponding author.\\ dong.phungvan@phenikaa-uni.edu.vn}
\affiliation{Phenikaa Institute for Advanced Study and Faculty of Basic Science, Phenikaa University, Yen Nghia, Ha Dong, Hanoi 100000, Vietnam}
\author{Duong Van Loi}
\email{loi.duongvan@phenikaa-uni.edu.vn}
\affiliation{Phenikaa Institute for Advanced Study and Faculty of Basic Science, Phenikaa University, Yen Nghia, Ha Dong, Hanoi 100000, Vietnam}

\begin{abstract}
The present matter content of our universe may be governed by a $U(1)_{B-L}$ symmetry---the simplest gauge completion of the seesaw mechanism which produces small neutrino masses. The matter parity results as a residual gauge symmetry, implying dark matter stability. The Higgs field that breaks the $B-L$ charge inflates the early universe successfully and then decays to right-handed neutrinos, which reheats the universe and generates both normal matter and dark matter manifestly.
\end{abstract}

\date{\today}
 
\maketitle

\section{Introduction}

The standard model must be extended in order to account for small neutrino masses and flavor mixing  \cite{Kajita:2016cak,McDonald:2016ixn} as well as dark matter component \cite{Jungman:1995df,Bertone:2004pz}. The seesaw mechanism is a compelling idea that realizes consistent neutrino masses, generated through the exchange of heavy right-handed neutrinos~\cite{Minkowski:1977sc,GellMann:1980vs,Yanagida:1979as,Glashow:1979nm,Mohapatra:1979ia,Mohapatra:1980yp,Lazarides:1980nt,Schechter:1980gr,Schechter:1981cv}. This mechanism can simply be realized in the gauge completion $U(1)_{B-L}$ for the standard model. Here, the right-handed neutrinos arise as a result of $B-L$ anomaly cancellation, while their heavy Majorana masses (or seesaw scale) are set by $B-L$ breaking scale. This $B-L$ dynamics also implies a natural leptogenesis, which generates the observed baryon asymmetry of the universe via CP-violating right-handed neutrino decay followed by sphaleron process \cite{Fukugita:1986hr,Buchmuller:2005eh,Davidson:2008bu}. 

Several analyses of the $U(1)_{B-L}$ model were presented in, e.g., \cite{Davidson:1978pm,Mohapatra:1980qe, Marshak:1979fm,Khalil:2006yi,Basso:2008iv,Accomando:2016sge} and its dark matter candidates were extensively signified by modifying the symmetry and/or particle content \cite{Montero:2007cd,Okada:2010wd,Montero:2011jk,Okada:2012sg,Basak:2013cga,Sanchez-Vega:2014rka,Rodejohann:2015lca,Okada:2019sbb}. As a matter of fact, a $Z_2$ symmetry was added to the above theory, protecting some candidate we want, for instance a right-handed neutrino or new neutral scalar, from decay. But let us ask what is the nature of the $Z_2$ symmetry and which is the correct/appropriate mechanism for producing the dark matter relic. In a recent research, Ma put forward that the $Z_2$ symmetry may originate from a lepton parity, recognizing the matter parity~\cite{Ma:2015xla}. But, like supersymmetry the matter parity must be imposed by hand, since the lepton number or R-symmetry is not conserved by the theory. Obviously, alternative choice for $Z_2$ (cf. \cite{Montero:2007cd,Okada:2010wd,Montero:2011jk,Okada:2012sg,Basak:2013cga,Sanchez-Vega:2014rka,Rodejohann:2015lca,Okada:2019sbb}) or matter parity (cf. \cite{Martin:1997ns,Dong:2006vk})\footnote{Including the cases that omit the extra symmetry while adding candidates with appropriate $B-L$ charges.} might lead to another solution for dark matter plus several unrelated production mechanisms, which are all due to the fact that we have not yet had an underlying principle governing dark matter physics. Our primary aim is to look for a more fundamental law in order to manifestly stabilize dark matter candidates and set their present abundance.  

In this paper, we prove that the $U(1)_{B-L}$ gauge symmetry for seesaw mechanism play such a law in regard to the dark matter sector. Indeed, this $U(1)_{B-L}$ theory can supply dark matter stability naturally after symmetry breaking, which recognizes the novel dark matter candidates, without requiring any ad hoc modification and extra symmetry\footnote{This interpretation is more elegant than the class of gauge theories recently studied in \cite{Dong:2013wca,Dong:2014wsa,Huong:2015dwa,Huong:2016ybt,Alves:2016fqe,Dong:2015yra,Dong:2015jxa,Dong:2016sat,Dong:2016gxl,Dong:2017zxo,Huong:2018ytz,Huong:2019vej,VanLoi:2019xud}.}. The dark matter relic is governed by just $B-L$ dynamics as produced through a leptogenesis, similar to that of the baryon asymmetry\footnote{See \cite{Nussinov:1985xr,Zurek:2013wia,Petraki:2013wwa,Falkowski:2011xh} for relevant discussions.}. Thus this new neutrino mass generation scheme also implies dark matter component and its abundance, besides the baryon asymmetry, which we will also prove that they all depart from the early universe inflation appropriately derived by the $B-L$ symmetry \cite{Starobinsky:1980te,Guth:1980zm,tanabashi:2018oca, Alexander:2004us, Buchmuller:2019gfy, Buchmuller:2013lra, Buchmuller:2012wn}.   

To be concrete, we reconsider the question of $B-L$ anomaly cancellation. We show that right-handed neutrinos can be divided into two kinds: (i) dark matter includes $N_R$ fields that have even $B-L$ number and (ii) normal matter contains $\nu_R$ fields that possess odd $B-L$ number. We prove that the matter parity arises naturally as a residual $B-L$ gauge symmetry, derived by a $B-L$ breaking scalar field. This scalar field inflates the early universe successfully and defines the seesaw scale. The fields $\nu_R$ obtain large Majorana masses in similarity to the often-studied right-handed neutrinos, which make observed neutrino masses small, whereas the fields $N_R$ have arbitrary masses providing a novel candidate for dark matter, stabilized by the matter parity conservation. We point out that the inflaton decays to a pair of $\nu_R$ or the Higgs field which reheats the early universe. As a result of the $U(1)_{B-L}$ gauge symmetry and matter parity, both the normal and dark matter abundances observed today are simultaneously generated by the CP-violating decays of the lightest $\nu_R$ in the early universe, analogous to the standard leptogenesis. Thus, this Abelian recognition of $B-L$ symmetry and matter parity is more simple than a previous proposal \cite{Dong:2018aak} and initiating a new research direction looking for dark matter candidates in connection to the baryon asymmetry production, where both kinds of the matter relics originate from the same source, addressed in a common framework.  

The rest of this work is arranged as follows. In Sec.~\ref{model}, we set up the model. In Sec.~\ref{scalar} we examine the potential minimization and scalar mass spectrum. In Sec.~\ref{neutrino} we discuss neutrino mass. In Sec. \ref{asymmetries} we obtain the dark and normal matter asymmetries. The other dark matter bounds are given in Sec. \ref{pheno}. Finally, we conclude this work in Sec. \ref{concl}. For completeness, in App.~\ref{inflation} we investigate cosmological inflation and reheating.       

\section{\label{model} The model}

The gauge symmetry is given by 
\be SU(3)_C\otimes SU(2)_L\otimes U(1)_Y\otimes U(1)_{B-L}, \ee where $B-L$ is baryon minus lepton charge, while the rest is the ordinary gauge group. The electric charge operator is related to the hypercharge by $Q=T_3+Y$, in which $T_i$ $(i=1,2,3)$ are $SU(2)_L$ weak isospin. 

The fermion content transfroms under the gauge symmetry as 
\bea 
&&Q_{aL}=(u_{aL}\ d_{aL})^T \sim (3,2,1/6,1/3),\\ 
&&u_{aR} \sim (3,1,2/3,1/3),\\ 
&& d_{aR}\sim (3,1,-1/3,1/3),\\
&&\psi_{aL}=(\nu_{aL}\ e_{aL})^T \sim (1,2,-1/2, -1),\\  && e_{aR}\sim (1,1,-1,-1),\\ 
&& \nu_{nR}\sim (1,1,0,x),\\ 
&& N_{mR}\sim (1,1,0,y).\eea Here $a=1,2,3$, $n=1,2,3,...,N$, and $m=1,2,3,...,M$ are family indices. The fields $\nu_R$ and $N_R$ are new fields, required in order to cancel $B-L$ anomalies, whereas the other fields define ordinary particles. Notice that the $B-L$ charge of the ordinary particles has been assigned analogous to the standard model, i.e. determined by the baryon or lepton number of them, motivated by the fact that the standard model and observed phenomena actually conserve $B-L$.

The nontrivial anomaly cancellation conditions are  
\bea [\mathrm{Gravity}]^2 U(1)_{B-L} &\sim&\sum_{\mathrm{fermions}}[(B-L)_{f_L}-(B-L)_{f_R}]\crn
&=&-(3+N x +M y)=0,\eea
\bea
[U(1)_{B-L}]^3 &=& \sum_{\mathrm{fermions}}[(B-L)_{f_L}^3-(B-L)_{f_R}^3]\crn
&=&-(3+Nx^3+My^3)=0.\eea 
The solutions with the smallest $M+N$ are $x=y=-1$ for $M+N=3$ and $(x,y)=(-4,5)$ for $(N,M)=(2,1)$, which were well-established in the literature, e.g., \cite{Costa:2019zzy,Montero:2007cd,Montero:2011jk,Sanchez-Vega:2014rka}. Such cases do not provide simultaneously dark matter candidates and successful leptogenesis. We consider the next solution for $N+M=4$, 
\be (x,y)=(-1,0)\ \mathrm{for}\ (N,M)=(3,1), \ee 
or in other words,
\be \nu_{1,2,3R}\sim (1,1,0,-1),\hs N_R\sim (1,1,0,0). \ee Here $N_R$ is a truly sterile neutrino under the gauge symmetry, which was actually omitted in the literature, e.g.,~\cite{Costa:2019zzy}. 

Besides the standard model Higgs doublet, \be \phi =(\phi^+\ \phi^0)^T\sim (1,2,1/2,0),\ee 
we introduce two scalar singlets, 
\be \varphi\sim (1,1,0,2),\hs \chi\sim (1,1,0, 1),\ee which are required to break the $B-L$ symmetry, giving new fermion masses, as well as supplying asymmetric dark matter. 

The minimal solution of $\nu_{aR},N_R$ and the minimal choice of $\varphi,\chi$ would yield viable phenomenological aspects, which are also strictly implied by the following residual gauge symmetry. Indeed, if one goes to the next step with $N+M=5$, a solution arisen is $(x,y)=(-1,1)$ with $(N,M)=(4,1)$. In this case, the new fermions do not supply dark matter stability naturally. Moreover, the solution with $(x,y)=(-1,0)$ for $(N,M)=(3,2)$ is not minimal. Similarly, while $\varphi,\chi$ are necessarily imposed, more scalar fields added would break the criteria of a minimal model.   

The Lagrangian is \be \mathcal{L}=\mathcal{L}_{\mathrm{kinetic}}+\mathcal{L}_{\mathrm{Yukawa}}-V,\ee where the first part defines kinetic terms and gauge interactions. Whereas, the Yukawa interactions and scalar potential are given, respectively, by  
\bea \mathcal{L}_{\mathrm{Yukawa}} &=& h^d_{ab}\bar{Q}_{aL}\phi d_{bR}+h^u_{ab} \bar{Q}_{aL}\tilde{\phi} u_{bR}\crn
&&+h^e_{ab}\bar{\psi}_{aL}\phi e_{bR}+h^\nu_{ab}\bar{\psi}_{aL}\tilde{\phi} \nu_{bR}\crn
&& +\fr 1 2 x_{ab}\bar{\nu}^c_{aR}\varphi\nu_{bR}+y_a \bar{\nu}^c_{aR}\chi N_R\crn
&& - \fr 1 2 m_{N}\bar{N}^c_R N_R+H.c.,\eea 
\bea V &=& \mu^2_1 \phi^\dagger \phi+\mu^2_2\varphi^* \varphi+\mu^2_3\chi^* \chi +[\mu \varphi^* \chi^2+H.c.]\crn
&&+\la_1 (\phi^\dagger \phi)^2+\la_2 (\varphi^* \varphi)^2+\la_3 (\chi^* \chi)^2\crn
&&+\la_4 (\phi^\dagger \phi)(\varphi^* \varphi)+\la_5 (\phi^\dagger \phi)(\chi^* \chi)\crn
&&+\la_6 (\varphi^* \varphi)(\chi^* \chi).\eea    
    
We can choose the potential parameters so that $\phi$ and $\varphi$ develop the vacuum expectation values (VEVs) such as
\be \langle \phi \rangle = \fr{1}{\sqrt{2}}(0\ v)^T,\hs \langle \varphi \rangle =\fr{1}{\sqrt{2}} \La,\ee while $\chi$ possesses vanishing VEV, i.e. $\langle \chi \rangle =0$.\footnote{Explicitly shown in the next section.} For consistency, one imposes \be \La \gg v=246\ \mathrm{GeV}.\ee 

The gauge symmetry is broken as\bc \begin{tabular}{c} $SU(3)_C\otimes SU(2)_L\otimes U(1)_Y\otimes U(1)_{B-L}$ \\
$\downarrow\La$\\
$SU(3)_C\otimes SU(2)_L\otimes U(1)_Y\otimes W_P$\\
$\downarrow v$\\
$SU(3)_C\otimes U(1)_Q\otimes W_P$ \end{tabular}\ec
where the first step implies the matter parity $W_P$, while the second step yields the electric charge $Q=T_3+Y$. 

The $B-L$ symmetry is broken by two units due to $[B-L](\varphi)=2$, but not complete, leading to a residual symmetry. Indeed, the matter parity conserves the VEV of $\varphi$, i.e. $W_P \La=\La$, where $W_P=e^{i\al (B-L)}$ is a $U(1)_{B-L}$ transformation. We obtain $e^{i2\al}=1$, implying $\al=k\pi$ for $k=0,\pm1,\pm2,\cdots$ Hence, $W_P=e^{ik\pi (B-L)}=[e^{i\pi (B-L)}]^k$. The field representations under $W_P$ is given in Table \ref{add1tb}. It is clear that $W_P=1$ for minimal $|k|=6$, except the identity with $k=0$. Therefore, $W_P$ is automorphic to $Z_6$ group, namely \[W_P=Z_6=\{1,p,p^2,p^3,p^4,p^5\}\] where $p\equiv e^{i\pi (B-L)}$ and $p^6=1$. We see that $Z_2=\{1,p^3\}$ is an invariant (normal) subgroup of $Z_6$. So, we factorize $W_P=Z_6=Z_2\otimes Z_3$, where $Z_3=Z_6/Z_2=\{Z_2,\{p,p^4\},\{p^2,p^5\}\}$ is the quotient group of $Z_6$ by $Z_2$. One further defines $Z_3=\{[1], [p^2],[p^4]\}$, where each coset element $[g]$ includes two elements of $Z_6$, the characteristic $g$ and the other $p^3g$ multiplied by $p^3$, and note that $[p^4]=[p^2]^2$, $[p^2]^3=1$. That said, $Z_2$ and $Z_3$ are generated by the generators $p^3=(-1)^{3(B-L)}$ and $[p^2]=[w^{3(B-L)}]$, respectively, where $w\equiv e^{i2\pi/3}$ is the cube root of unity. Notice that since $p^6=e^{i\pi 6 (B-L)}=1$, we have $3(B-L)$ to be integer. Hence $p^3=1$ or $-1$ correspond to the representations $\underline{1}$ or $\underline{1}'$ of $Z_2$, while $[p^2]=[1]\rightarrow 1$, $[w]\rightarrow w$, or $[w^2]\rightarrow w^2$ correspond to the representations $\underline{1}$, $\underline{1}'$, or $\underline{1}''$ of $Z_3$, where the isomorphism ``$\rightarrow$'' defines the corresponding representation. The field representations under component subgroups $Z_2,Z_3$ are also added to Table \ref{add1tb}. 
\begin{table}[h]
\bc
\begin{tabular}{c|cccccc}
\hline\hline
Field & $(\nu,e)$ & $(u,d)$ & (gauge boson, $\phi$) & $\varphi$ & $N$ & $\chi$\\
\hline
$W_P$ & $(-1)^k$ & $e^{ik\pi/3}$ & 1 & 1 & 1 & $(-1)^k$\\ 
\hline
$p^3$ & $-1$ & $-1$ & 1 & 1 & 1 & $-1$\\ 
\hline
$[p^2]$ & $1$ & $w$ & $1$ & $1$ & $1$ & $1$\\ 
\hline\hline
\end{tabular}
\caption[]{\label{add1tb} Field representations under the residual symmetry, where the $W_P$ value of a fermion does not depend on its left or right chirality and generation index.}
\ec
\end{table} 

It is clear that the only quarks transform nontrivially under the quotient group $Z_3$. Therefore, the theory automatically conserves $Z_3$ because of $SU(3)_C$ symmetry; in other words, $Z_3$ is accidentally conserved by $SU(3)_C$. Neglecting the quotient group $Z_3$, the residual symmetry $W_P$ is already defined by $Z_2=\{1,p^3\}$.  
In other words, considering the residual symmetry for $k=3$, we get $W_P=(-1)^{3(B-L)}$. The matter parity is conveniently rewritten as \be W_P=(-1)^{3(B-L)+2s}\ee after multiplying the spin parity $(-1)^{2s}$, which is conserved by the Lorentz symmetry. 

At this step, $W_P$ is odd, if \[B-L=\fr{1+2k}{3}=\pm\fr 1 3,\pm 1,\pm \fr 5 3,\cdots\] for boson, and 
\[B-L=\fr{2k}{3}=0,\pm\fr 2 3,\pm \fr 4 3,\cdots\] for fermion. In view of $W_P$, the field $N_R$ that has $B-L=0$ is the minimal solution for dark fermions, while the field $\chi$ that has $B-L=1$ (where the conjugated $\chi$ has $B-L=-1$) is the minimal solution for dark scalars. Here notice that the bosonic solution for $B-L=\pm 1/3$ was not interpreted, since it disturbs the quotient $Z_3$ group, not ensured by $SU(3)_C$.

Moreover, the notation ``$W$'' means particles that have ``wrong'' $B-L$ number and odd under the matter parity (i.e. $W_P=-1$), say $N_R$ and $\chi$, called wrong particles.\footnote{Namely, they have basic $B-L$ charge---not counting for charge addition due to the cyclic property of residual symmetry---different from the standard model definition.} All the other particles, including the standard model, $\nu_R$, $\varphi$, and $U(1)_{B-L}$ gauge (called $Z'$) fields, are even under the matter parity (i.e. $W_P=1$), which have normal $B-L$ number or differ from that number by even unit as $\varphi$ does, called normal particles. They are summarized in Table~\ref{tb1}. 
\begin{table}[h]
\bc
\begin{tabular}{c|ccccccccccccc}
\hline\hline
Particle & $\nu$ & $e$ & $u$ & $d$ & gluons & photon & $W$ & $Z$ & $Z'$ & $\phi$ & $\varphi$ & $N$ & $\chi$\\
\hline
$W_P$ & 1 & 1 & 1 & 1 & 1& 1& 1& 1& 1& 1 & 1 & $-1$ & $-1$ \\
\hline\hline
\end{tabular}
\caption[]{\label{tb1} Matter parity for the model particles}
\ec
\end{table}

It is easily realized that the $\chi$ vacuum value vanishes, $\langle \chi \rangle =0$, due to the matter parity conservation. Also, the lightest wrong particle (LWP) between $N_R$ and $\chi$ is absolutely stabilized responsible for dark matter. Furthermore, the new observation is that $\nu_R$ couples both $N \chi$ and $e\phi$, through the complex Yukawa couplings, $y$ and $h^\nu$, respectively. Hence, the asymmetric dark and normal matter can be simultaneously produced by CP-violating decays of $\nu_R$, in the same manner of the standard leptogenesis. Of course, the $\nu_R$ fields are generated after cosmic inflation derived by the inflaton $\varphi$---the scalar field that breaks $B-L$---which also induces the neutrino seesaw masses. Let us see.       

\section{\label{scalar} Scalar potential}

The scalar potential implies the gauge symmetry breaking. First, the $\varphi$ field obtains a large VEV derived by $V(\varphi)=\mu^2_2\varphi^*\varphi +\la_2 (\varphi^*\varphi)^2$ to be \be \La^2=-\mu^2_2/\la_2,\ee provided that $\la_2>0$, $\mu^2_2<0$, and $|\mu_2|\gg |\mu_{1,3}|$. 

Integrating $\varphi$ out, one finds that the effective potential at leading order as \bea V(\phi,\chi)&=&\mu^2_1\phi^\dagger \phi +\mu^2_3\chi^*\chi +\la_1(\phi^\dagger \phi)^2+ \la_3 (\chi^*\chi)^2\crn
&&+\la_5(\phi^\dagger \phi)(\chi^*\chi).\eea Note that the mixing terms $(\mu \varphi^*\chi^2+H.c.)+\varphi^*\varphi(\la_4 \phi^\dagger \phi+\la_6\chi^*\chi)$ between $\varphi$ and $(\phi,\chi)$ give small contributions, assuming that $|\la_4|\ll |\mu^2_1|/\La^2$, $|\la_6|\ll |\mu^2_3|/\La^2$, and $|\mu|\ll |\mu^2_3|/\La$, such that the two sectors, $\varphi$ and $(\phi,\chi)$, are approximately decoupled. 

Here, a question arisen is that the $\varphi$ contribution to the standard model Higgs and dark scalar masses, such as $\fr 1 2 \La^2 (\la_4 \phi^\dagger \phi+\la_6 \chi^*\chi)$, create a hierarchy, instabilizing the electroweak vacuum? This problem can be cured by setting $\la_4=0=\la_6$ at the tree level as well as imposing a classical conformal symmetry that suppresses tree-level mass parameters, analogous to \cite{Iso:2009nw,Iso:2009ss,Das:2015nwk}. Then $\la_{4,6}$ are induced at the one-loop level by new fermion contributions and at the two-loop level by top quark, similar to the diagrams shown in \cite{Iso:2009nw,Iso:2009ss,Das:2015nwk}. The induced coupling $\la_4$ can be so small and negative which triggers the electroweak symmetry breaking, whereas the induced coupling $\la_6$ can be so small and positive which conserves the matter parity. Such couplings transmute to the expected, finite Higgs and dark-scalar masses. The hierarchy question in this framework is worth exploring to be published elsewhere. 

Choosing the parameters as $\mu^2_1<0$, $\mu^2_3>0$, $\la_{1,3,5}>0$ we derive the VEVs from $V(\phi,\chi)$ to be
\be v^2=-\mu^2_1/\la_1,\hs \langle \chi \rangle=0. \ee    

The physical scalar fields with corresponding masses are given as 
\bea && \varphi = \fr{1}{\sqrt{2}}\left(\La +H' + i G_{Z'}\right),\hs m^2_{H'}=2\la_2 \La^2, \\
&& \phi = \left(\begin{array}{c}
G^+_W\\ \fr{1}{\sqrt{2}}(v+H+i G_Z)\end{array}\right),\hs m^2_{H}=2\la_1 v^2, \\
&& \chi, \hs m^2_\chi=\mu^2_3. \eea Here $H$ is identical to the standard model Higgs boson, while $H'$ is a new heavy Higgs boson associate to $B-L$ symmetry breaking. $G_W$, $G_Z$, and $G_{Z'}$ are massless Goldstone bosons eaten by $W$, $Z$, and $Z'$ gauge bosons, respectively. $\chi$ has an arbitrary mass $m_\chi$, but below the $\La$ scale.

\section{\label{neutrino} Neutrino mass} 

The ordinary fermions obtain appropriate masses as in the standard model. The new fermions get masses as follows 
\bea \mathcal{L}_{\mathrm{Yukawa}} &\supset& -\fr 1 2 (\bar{\nu}_L \bar{\nu}^c_R)\left(\begin{array}{cc}
0 & m_D \\
m^T_D & m_R\end{array}\right)\left(\begin{array}{c}
\nu^c_L \\ \nu_R \end {array}\right)\crn
&&-\fr 1 2 m_N \bar{N}^c_{R} N_R+H.c., \eea where we define $\nu = (\nu_{1}\ \nu_{2}\ \nu_{3})^T$ and
\be [m_D]_{ab}=-h^\nu_{ab}\fr{v}{\sqrt{2}},\hs [m_R]_{ab}=-x_{ab}\fr{\La}{\sqrt{2}}.\label{adddt12}\ee

The dark fermion $N_R$ gets an arbitrary mass $m_N$. The observed neutrinos $(\sim \nu_L)$ gain a mass via the seesaw mechanism due to $v\ll \La$ to be 
\be m_\nu\simeq -m_D m^{-1}_R m^T_D=h^\nu x^{-1}(h^\nu)^T\fr{v^2}{\sqrt{2}\La}.\label{adddt1212}\ee Comparing to the neutrino data, $m_\nu\sim 0.1$ eV, we get \be \La \sim [(h^\nu)^2/2x] \times 10^{15}\ \mathrm{GeV}\sim 10^{15}\ \mathrm{GeV},\ee which is naturally at the inflation scale and is hereafter taken into account. Of course, the heavy neutrinos $\sim \nu_R$ have the mass, $m_R$, proportional to the $\La$ scale.

The cosmic inflation can be successfully derived by the inflaton field $\varphi$ when it nonminimally interacts with gravity, analogous to the Higgs inflation \cite{Bezrukov:2007ep}. This is briefly presented in Appendix \ref{inflation}; additionally, the interested reader can look at \cite{Oda:2017zul} for more detailed treatment. Also from Appendix \ref{inflation} for estimation of the $x$ coupling, we obtain the right-handed neutrino mass scales, \be m_{\nu_{1R}}\sim 10^{11}\ \mathrm{GeV},\hs m_{\nu_{2,3R}}\gtrsim 10^{13}\ \mathrm{GeV}.\label{dtttn1}\ee 

With the mass $m_{\nu_{1R}}< 10^{11}$ GeV, the reheating happens immediately after the end of inflation yielding a reheating temperature $T_R\sim 4.4\times 10^{10}$ GeV, where $\nu_{1R}$ is present in the thermal bath of the universe. However, with the mass $m_{\nu_{1R}}>10^{11}$ GeV there is a period of preheating, waiting for necessary oscillations of inflaton before decay. In this case, the nonperturbative decay productions of inflaton to $Z'Z'$ can rapidly thermalize, making a cosmic plasma with background temperature much higher than the reheating temperature, e.g. $10^3T_R$ \cite{Chung:1998rq}. Consequently, all the right-handed neutrinos can be created through thermalizing of $Z'$ or lighter states during the preheating. Hence, in the following we mainly consider thermal lepogenesis scenario coming from $\nu_{1R}$ decay.

\section{\label{asymmetries} Asymmetric matter}

The Yukawa Lagrangian yields 
\bea \mathcal{L} &\supset& -\bar{e}_{aL}[m_e]_{ab}e_{bR}-\fr 1 2 m_N \bar{N}^c_R N_R\crn
&&-\fr 1 2 \bar{\nu}_{aL}[m_\nu]_{ab}\nu^c_{bL}-\fr 1 2 \bar{\nu}^c_{aR}[m_R]_{ab}\nu_{bR}\crn
&&+h^\nu_{ab}\bar{\psi}_{aL}\tilde{\phi}\nu_{bR}+y_b \bar{N}^c_R\chi\nu_{bR}+H.c.,\label{eqd1}\eea where $m_e\equiv -h^e v/\sqrt{2}$, while $m_\nu$ and $m_R$ were given in (\ref{adddt1212}) and (\ref{adddt12}), respectively. The mixing effect (mixing angle) between $\nu_L$ and $\nu_R$ is too small, i.e. $\Theta\simeq m_D m^{-1}_R\sim v/\La\ll 1$, as omitted. 

The gauge states $(_a)$ are related to the mass eigenstates, subscripted by $(_i)$ for $i=1,2,3$, through mixing matrices, \be e_{aL,R}=[V_{eL,R}]_{ai}e_{iL,R},\hs \nu_{aL,R}=[V_{\nu L,R}]_{ai}\nu_{iL,R},\ee such that the corresponding mass matrices are diagonalized, \bea V^\dagger_{eL} m_e V_{eR} &=& \mathrm{diag}(m_e,m_\mu,m_\tau),\\
V^\dagger_{\nu L} m_\nu V^*_{\nu L} &=& \mathrm{diag}(m_{1},m_{2},m_{3}),\\ 
V^T_{\nu R} m_R V_{\nu R} &=& \mathrm{diag}(m_{\nu_{1R}},m_{\nu_{2R}},m_{\nu_{3R}}).\eea Here $m_{1,2,3}$ are the active neutrino masses (as observed), whereas $m_{\nu_{1,2,3R}}$ are the sterile neutrino masses. Concerning the sterile neutrino sector we assume a hierarchical $m_{\nu_{1R}}< m_{\nu_{2,3R}}$ and flavor-diagonal $V_{\nu R}=1$ without loss of generality. The last equality means that $\nu_{aR}=\nu_{iR}$ for $a=i=1,2,3$ are physical Majorana fields by themselves.     

We rewrite (\ref{eqd1}) in the mass bases,
\bea \mathcal{L}&\supset& -m_{e_i}\bar{e}_i e_i-\fr 1 2 m_N \bar{N}^c_R N_R\crn
&&-\fr 1 2 m_i \bar{\nu}_{iL}\nu^c_{iL}-\fr 1 2 m_{\nu_{iR}} \bar{\nu}^c_{iR}\nu_{iR}\crn
&&+z_{ij} \bar{\psi}_{iL}\tilde{\phi}\nu_{jR}+y_{j}\bar{N}^c_R\chi \nu_{jR}+H.c.,\eea where all $i,j$ indices are summed, and $\psi_{iL}=(\nu'_{i L}\ e_{iL})^T$ with $\nu'_{iL}=[V^\dagger_{eL}V_{\nu L}]_{ij}\nu_{jL}$ related through the lepton mixing matrix. The Yukawa couplings $z_{ij}=[V^\dagger_{eL} h^\nu]_{ij} $ and $y_j$ are all complex, hence they are the sources of CP violation. 

\begin{figure}[!h]
\bc
\includegraphics[scale=0.8]{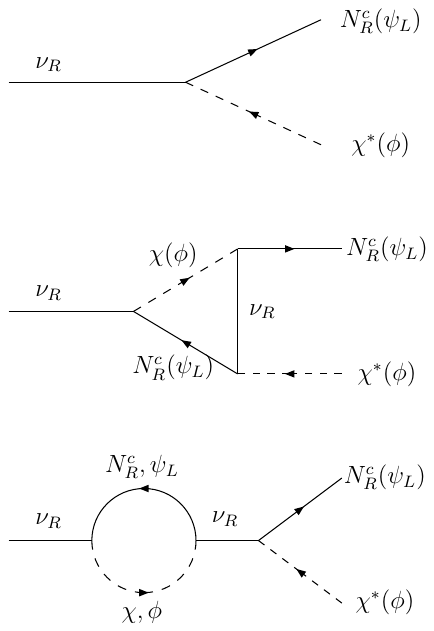}
\caption{\label{fig1} CP-violating decays of $\nu_R$ that produce both dark matter $N\chi$ and normal matter $(\psi\phi)$, respectively.} 
\ec
\end{figure}

It is noteworthy that in the early universe, the lightest right-handed neutrino $\nu_{1R}$ decays simultaneously to normal matter, $\nu_{1R}\rightarrow \psi_i \phi$, and dark matter, $\nu_{1R}\rightarrow N \chi$, through the diagrams, supplied in Fig. \ref{fig1}, up to one-loop level. This produces the corresponding CP asymmetries, responsible for observed matter relics, defined by
\bea \epsilon^i_{\mathrm{NM}}&=&\fr{\Ga(\nu_{1R}\rightarrow \psi_i \phi)-\Ga(\nu_{1R}\rightarrow \bar{\psi}_i \bar{\phi})}{\Ga_{\nu_{1R}}},\crn 
\epsilon_{\mathrm{DM}}&=&\fr{\Ga(\nu_{1R}\rightarrow N \chi)-\Ga(\nu_{1R}\rightarrow \bar{N} \bar{\chi})}{\Ga_{\nu_{1R}}},\eea where $\Ga_{\nu_{1R}}=m_{\nu_{1R}}(2[z^\dagger z]_{11}+y_1^* y_1)/16\pi$ is the total width of $\nu_{1R}$. 

Notice that the flavor effect can be important during the electroweak sphaleron since the interaction rates of charged lepton flavors through the Yukawa couplings to the standard model Higgs boson, i.e. $\Ga_l \simeq 5\times 10^{-3}(h^l)^2 T$ for $l=e,\mu,\tau$, may wash out generated asymmetries in $\nu_{1R}$ decay \cite{Nardi:2006fx,Cline:1993bd}. However, at the temperature of asymmetric $\nu_{1R}$ decay, $T=m_{\nu_{1R}}\simeq 10^{11}$ GeV, the Hubble rate $H_1=0.33\sqrt{g_*}T^2/m_P\simeq 1.37\times 10^{4} $ GeV is nearly comparable to the largest rate of tau flavor $\Ga_\tau\simeq 5\times 10^{4}$ GeV. The flavor separation if presented is only between the tau flavor and the combined state of muon and electron flavors, which as explicitly shown in \cite{Dong:2018aak} is smaller than the single flavor contribution. For simplicity, the lepton asymmetry production will be considered to be independent of flavor effect, yielding a net contribution by summing $\epsilon_{\mathrm{NM}} = \sum_i \epsilon^i_{\mathrm{NM}}$. Applying the Feynman rules for the diagrams in Fig. \ref{fig1} we obtain
\bea \epsilon_{\mathrm{NM}} &=& \fr{1}{8\pi(2[z^\dagger z]_{11}+y^*_1y_1)}\sum_{j}\Im\left\{(3[z^\dagger z]_{j1}+y^*_jy_1)[z^\dagger z]_{j1}\right\}r_{1j},\\ 
\epsilon_{\mathrm{DM}} &=& \fr{1}{8\pi(2[z^\dagger z]_{11}+ y^*_1y_1)}\sum_{j}\Im\left[([z^\dagger z]_{j1}+y^*_j y_1)(y^*_j y_1)\right]r_{1j}, \eea where $r_{1j}=m_{\nu_{1R}}/m_{\nu_{jR}}$, in agreement with \cite{Falkowski:2011xh}. 

The abundance yield of a particle $\al$ is defined as $Y_\al = n_\al/s$, where $n_\al$ and $s$ are the $\al$ particle number density and the entropy density, respectively. The abundance yield of a particle asymmetry is $Y_{\Delta \al}=Y_{\al}-Y_{\bar{\al}}$, which should vanish at early times. The evolution of the sterile neutrino $Y_{\nu_{1R}}$ and of the normal $Y_{\Delta \psi}$ and dark $Y_{\Delta N}$ matter asymmetries are determined by the Boltzmann equations. Like the Higgs doublet $\phi$, the asymmetry of $\chi$ is rapidly erased due to the fast interactions, e.g. the $\la_5$ coupling, that convert $\chi\leftrightarrow \bar{\chi}$. Thus, the asymmetries of $\phi,\chi$ would vanish, which are not considered. The Boltzmann equations include $\nu_{1R}$ decays, inverse decays, and 2-to-2 scatterings between the normal ($\psi\phi$) and dark $(N\chi)$ sectors, such as 
\bea && \fr{s H_1}{w} \fr{dY_{\nu_{1R}}}{dw}=-\ga_D\left(\fr{Y_{\nu_{1R}}}{Y^{\mathrm{eq}}_{\nu_{1R}}}-1\right),\label{them0} \\
&& \fr{s H_1}{w} \fr{dY_{\Delta \psi}}{dw}=\ga_D\left[\ep_{\mathrm{NM}}\left(\fr{Y_{\nu_{1R}}}{Y^{\mathrm{eq}}_{\nu_{1R}}}-1\right)-\fr{Y_{\Delta \psi}}{2Y^{\mathrm{eq}}_{\psi}}\mathrm{Br_\psi}\right]+\cdots, \label{them1}\\
&& \fr{s H_1}{w} \fr{dY_{\Delta N}}{dw}=\ga_D\left[\ep_{\mathrm{DM}}\left(\fr{Y_{\nu_{1R}}}{Y^{\mathrm{eq}}_{\nu_{1R}}}-1\right)-\fr{Y_{\Delta N}}{2Y^{\mathrm{eq}}_{N}}\mathrm{Br}_N\right]+\cdots,\label{them2} \eea where $w=m_{\nu_{1R}}/T$, $\ga_D=[m^3_{\nu_{1R}}K_1(w)/\pi^2 w]\Ga_{\nu_{1R}}$ is the thermally averaged $\nu_{1R}$ decay density, $\mathrm{Br}_{\psi,N}=\Ga_{\psi,N}/\Ga_{\nu_{1R}}$ are branching ratios that relate to the decay rates, $\Ga_{\psi}=m_{\nu_{1R}}[z^\dagger z]_{11}/8\pi$ and $\Ga_N=m_{\nu_{1R}}y^*_1 y_1/16\pi$, of $\nu_{1R}$ to $\psi \phi$ and $N\chi$ respectively, and the dots denote the mentioned scatterings. The terms that include $\ep_{\mathrm{NM,DM}}$ create $Y_{\Delta \psi,\Delta N}$ asymmetries when $\nu_{1R}$ drops out of equilibrium, while the terms of $\mathrm{Br}_{\psi,N}$ present the inverse decays that wash out the generated asymmetries. 

We are in the narrow-width approximation, i.e. $\Ga_{\nu_{1R}}\ll m_{\nu_{1R}}$ and $\Ga^2_{\nu_{1R}}/H_1\ll m_{\nu_{1R}}$, where $\Ga_{\nu_{1R}}/H_1$ is not too small so that $\nu_{1R}$ decays before dominating the universe. The inverse decays are the main source of washout, whereas the 2-to-2 processes can be neglected, as mediated by $\nu_{1R}$ and $Z'$ to be much small. The equations (\ref{them1}) and (\ref{them2}) are decoupled from each other, hence the corresponding asymmetries evolve independently, parameterized as 
\be Y_{\Delta \psi}= \eta_\psi \ep_{\mathrm{NM}} Y^{\mathrm{eq}}_{\nu_{1R}}(0),\hs Y_{\Delta N}= \eta_N \ep_{\mathrm{DM}} Y^{\mathrm{eq}}_{\nu_{1R}}(0), \ee where $Y^{\mathrm{eq}}_{\nu_{1R}}(0)=135\zeta(3)/(4\pi^4g_*)\simeq 4\times 10^{-3}$. The efficiency factors $\eta_{\psi,N}$ are obtained by solving the Boltzmann equations, respectively. It is shown that the normal sector with seesaw masses typically lies in the strong-washout regime, leading to an approximate solution, $\eta_\psi\simeq H_1/\Ga_{\psi}\ll 1$ \cite{Falkowski:2011xh}. The washout in the dark sector is considered suitably to be weak, $\Ga_{N}/H_1\ll 1$, yielding the solution, $\eta_N\simeq 1$.\footnote{The washout effect in the dark sector can be moderate or strong but is not interested in this work.} 
Additionally, the sphaleron process converts only the lepton asymmetry to baryon asymmetry, given by $Y_B=(12/37)Y_{\Delta \psi}$ similar to the standard model, while it does not convert the dark matter asymmetry to baryon due to the matter parity conservation. The observational data implies $\Om_{\mathrm{DM}}\simeq 5 \Om_B$ \cite{tanabashi:2018oca}, which leads to  
\be \fr{m_{N}}{m_{p}}\simeq \fr{5Y_B}{Y_{\Delta N}}\simeq \fr 5 3 \eta_\psi \fr{\ep_{\mathrm{NM}}}{\ep_{\mathrm{DM}}},\ee where $m_p$ and $m_N$ are the proton and fermion dark matter masses, respectively. Further, we use the experimental measurement of baryon asymmetry normalized to photon number $5.8\times 10^{-10}\leq n_B/n_\ga\leq 6.5\times 10^{-10}$ at $95\%$ CL, which correspondingly translates to \be Y_B\simeq 1.3\times 10^{-3} \eta_\psi \ep_{\mathrm{NM}} = (0.82\mathrm{-}0.92) \times 10^{-10},\label{cmbba1}\ee with the aid of $n_\ga/s\simeq 0.142$ \cite{tanabashi:2018oca}.

Following \cite{Casas:2001sr}, the Dirac Yukawa coupling of neutrinos satisfies 
\be [z^\dagger z]_{ij}=(2/v^2)\sqrt{m_{\nu_{iR}}m_{\nu_{jR}}}[R^\dagger.\mathrm{diag}(m_1,m_2,m_3).R]_{ij},\ee where $R$ is an arbitrary orthogonal matrix parameterized in terms of three Euler (complex) angles, $\hat{\theta}_{1,2,3}$, such as
\be R=\left(\begin{array}{ccc}
\hat{c}_2 \hat{c}_3 & -\hat{c}_1 \hat{s}_3 - \hat{s}_1 \hat{s}_2 \hat{c}_3 & \hat{s}_1 \hat{s}_3 - \hat{c}_1 \hat{s}_2 \hat{c}_3 \\
\hat{c}_2 \hat{s}_3 & \hat{c}_1 \hat{c}_3 - \hat{s}_1 \hat{s}_2 \hat{s}_3 & -\hat{s}_1 \hat{c}_3 - \hat{c}_1 \hat{s}_2 \hat{s}_3\\
\hat{s}_2 & \hat{s}_1 \hat{c}_2 & \hat{c}_1 \hat{c}_2
\end{array}\right),\ee where $\hat{c}_i=\cos \hat{\theta}_i$ and $\hat{s}_i=\sin{\hat{\theta}}_i$. For numerical investigation, we take $m_{\nu_{1R}}=10^{11}$ GeV and $m_{\nu_{2R}}=m_{\nu_{3R}}=10^{13}$~GeV, as implied from (\ref{dtttn1}). We consider the active neutrinos having a normal hierarchical spectrum, such as $m_1=0$, $m_2=8.6\times 10^{-3}$ eV, and $m_3=0.05$ eV \cite{tanabashi:2018oca}. Further, we assume $y_3=y_2=y_1 e^{-i\theta}$, where $y_1$ is real since its phase can be removed by redefining appropriate fields in the relevant Yukawa interaction. With the given value of $m_{\nu_{1R}}$, the weak washout present in the dark sector implies $\Ga_N/H_1\simeq (3.8\times 10^{2}y_1)^2\ll1$. So we fix $y_1=10^{-3}$. To ascertain the strong washout in the normal sector, we compute $\Ga_\psi/H_1\simeq 10^3[R^\dagger m_\nu R]_{11}/\mathrm{eV}=8.6|\hat{c}_2|^2|\hat{s}_3|^2+50|\hat{s}_2|^2\gg 1$, thus either $\hat{\theta}_2$ or $\hat{\theta}_3$ is finite. Therefore, we put $\hat{\theta}_3=\pi/2$. 

\begin{figure}[!h]
\bc
\includegraphics[scale=0.5]{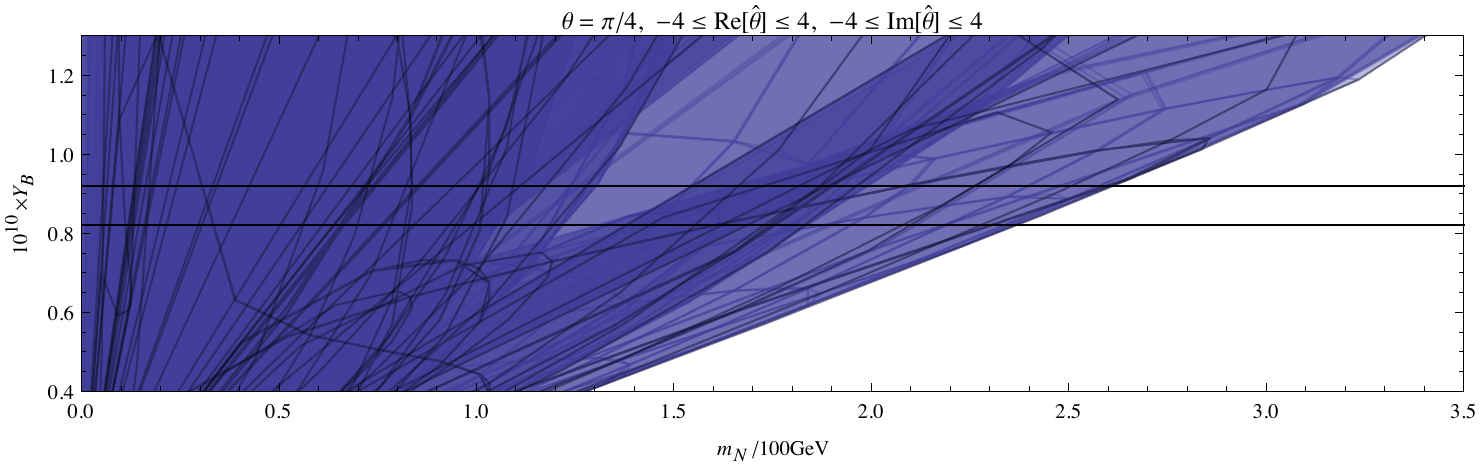}\\
\vs
\includegraphics[scale=0.56]{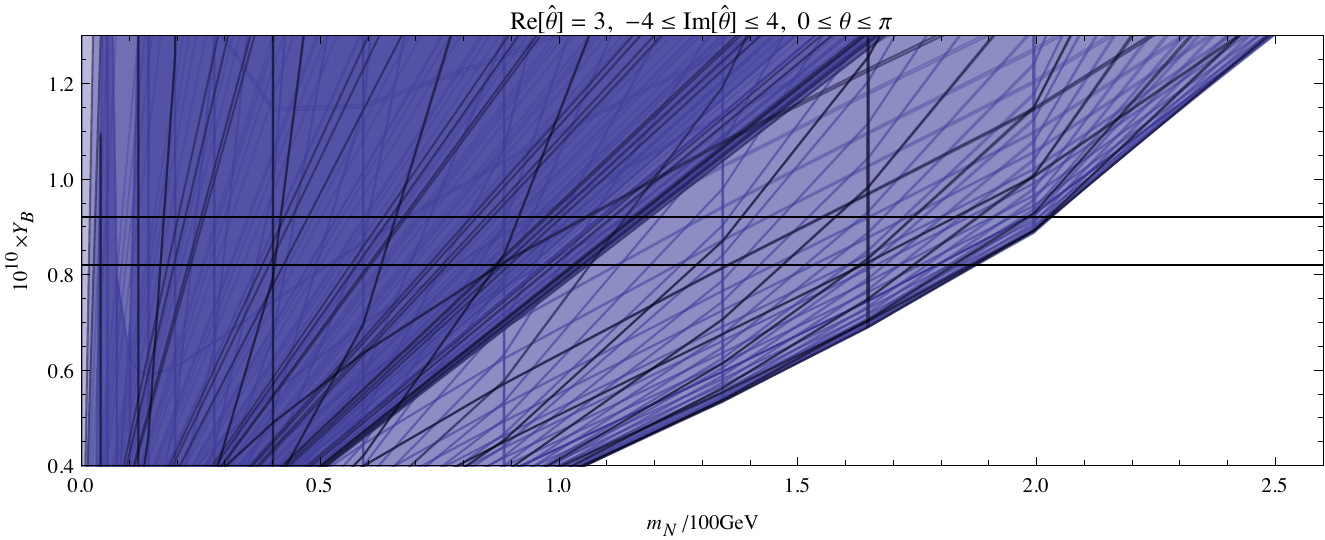}\\
\vs
\includegraphics[scale=0.52]{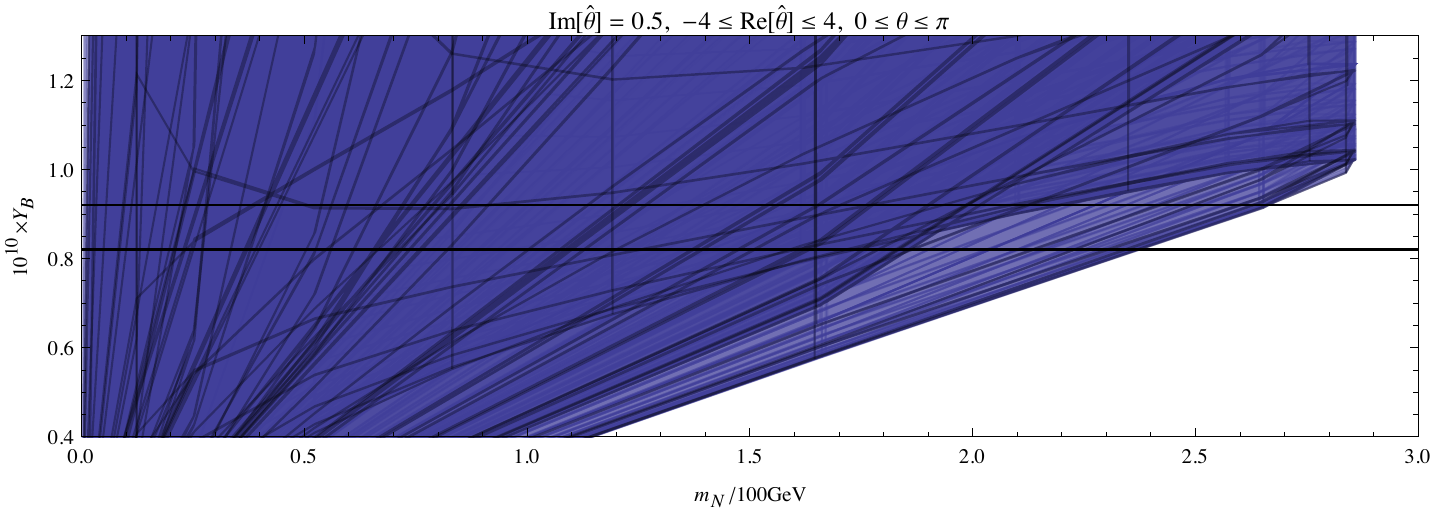}
\caption{\label{fig2} Baryon asymmetry $Y_B$ and fermion dark-matter mass $m_N$ despited as the function of two CP-violating parameters determined in the ranges as shown on the top of each panel whereas the first CP-parameter is fixed.} 
\ec
\end{figure}

We plot the observable quantities $(Y_B,m_N/m_p)$ on a coordinate plane as the function of parameters, $\hat{\theta}_1=\hat{\theta}_2\equiv \hat{\theta}$ and $\theta$, in the regions, $-4<\Re(\hat{\theta})<4$, $-4<\Im(\hat{\theta})<4$ and $0<\theta<\pi$, respectively. Having totally three parameters, one of them would be fixed, while the other two vary, yielding three figures as indicated in Fig. \ref{fig2}. There, we also show the $Y_B$ bounds as the horizontal lines in each panel. With the choice of the parameters, the fermion mass is limited by $m_N\sim 250$ GeV, much larger than the proton mass. Although the $N$ dark matter is asymmetrically produced in the thermal bath of the universe, its mass may be below the electron mass---the decoupling mass limit of thermal relics. Hence, $N$ can take a mass between the keV and weak scales.       

Specially, we may have a scenario of nonthermal leptogenesis, given that $\nu_{1R}$ is directly produced by the inflaton decay $\Phi\rightarrow \nu_{1R}\nu_{1R}$. In this case, the reheating temperature may be lower than the lightest right-handed neutrino mass. The total lepton asymmetry is simply summed of flavor asymmetries. The lepton and dark matter asymmetry densities normalized to photon density relate to the CP asymmetries by \be \eta_{\mathrm{NM},\mathrm{DM}}=\fr 3 2 \epsilon_{\mathrm{NM,DM}}\times \mathrm{Br}(\Phi\rightarrow \nu_{1R}\nu_{1R})\times \fr{T_R}{m_\Phi}.\ee This leads to $\eta_{\mathrm{NM}}/\eta_{\mathrm{DM}}=\epsilon_{\mathrm{NM}}/\epsilon_{\mathrm{DM}}$. The remark is that this nonthermal scenario produces asymmetries coinciding with the thermal scenario above when thermally produced $\nu_{1R}$ with washout effects in both normal and dark sectors to be weak, i.e. $\eta_{\psi}\simeq 1\simeq \eta_N$. Indeed, the latter case implies $\eta_{\mathrm{NM}}/\eta_{\mathrm{DM}}\simeq \epsilon_{\mathrm{NM}}/\epsilon_{\mathrm{DM}}$ as approximately hold as the former case, where the thermal leptogenesis is flavor independent and very effective. All the cases imply   
\be \fr{m_{\mathrm{DM}}}{m_{p}}\simeq \fr{5\eta_B}{\eta_{\mathrm{DM}}}\simeq \fr{5}{3} \fr{\epsilon_{\mathrm{NM}}}{\epsilon_{\mathrm{DM}}}.\ee
It is evident from (\ref{cmbba1}) that $\ep_{\mathrm{NM}}\simeq 10^{-7}$ for $\eta_\psi\simeq 1$. Assuming $\Ga_N\gg \Ga_\psi$, i.e. $y_1\gg 10^{-2}(|\hat{s}_2|,|\hat{c}_2 \hat{s}_3|)$, and taking $m_{\nu_{2R}}=m_{\nu_{3R}}=10^2 m_{\nu_{1R}}$ and $y_2=y_3=y_1 e^{-i\theta}$ with real $y_1$, as before, we derive 
\be  \fr{m_{\mathrm{DM}}}{m_{p}}\sim  \left(\fr{10^{-2}}{y_1}\right)^2 \fr{1}{\sin(2\theta)}.\label{dtttn12}\ee Now, the condition $\ep_{\mathrm{NM}}\simeq 10^{-7}$ yields $|\hat{c}_2|^2|\hat{s}_1|\sin(\theta-\mathrm{Arg}\hat{s}_1)\sim 10^{-2}$. Provided that $y_1\sim 10^{-2}$ and $\sin{2\theta}\sim 10^{-3}$--1, from (\ref{dtttn12}) we have $m_{\mathrm{DM}}\sim 1$--1000 GeV. A smaller mass of dark matter can be obtained if $y_1$ is bigger than $10^{-2}$.  

\section{\label{pheno} Dark matter detection}

In this section, we will examine the direct detection experiments of dark matter and the dark matter signatures at colliders. Simultaneously, we will outline the multicomponent nature of dark matter that the model owns and indicates oppotunities to probe them.   

\subsection{Dark-matter nucleon scattering}

As obtained in the previous section, the thermal leptogenesis yields a fermion dark matter candidate, $N$. Its present abundance is asymmetrically produced through the CP-violating decay of the lightest right-handed neutrino which is out of equilibrium. The $N$ mass should be smaller than that of the dark scalar $\chi$, so that $N$ is stabilized by the matter parity. Whereas, the heavier candidate $\chi$ has the vanishing present density due to the fast interaction with the standard model Higgs boson $H$ as well as completely annihilating to the normal particles through such Higgs portal. Additionally, the heavy state $\chi$ if created by some source\footnote{Such as particle colliders or inelastic scattering of $N$ dark matter with thermal neutrinos in the sun.} can decay to the stable light state $N$ plus normal fields, through the coupling $y_j$. Since $\nu_{jR}$ is superheavy, the decay only proceeds due to the mixing of $\nu_{jR}$ with the standard model neutrinos. That said, one has a coupling $y_j \Theta_{jk} \bar{N}^c_R\chi \nu^c_{kL}$, where $\Theta_{jk}\sim v/\La$ is the $\nu_L$--$\nu_R$ mixing matrix element in the seesaw mechanism, as mentioned. The decay rate is \bea \Ga(\chi\rightarrow N^c_R \nu_L) &\simeq & \fr{m_\chi}{8\pi} \sum_k y^2_j \Theta^2_{jk} \sim \left(\fr{m_\chi}{100\ \mathrm{GeV}} \right) \left(\fr{y_j}{10^{-3}} \right)^2 \left(\fr{10^{13} v}{\La} \right)^2 4\times 10^{-32}\ \mathrm{GeV}.\eea With the benchmark values of the parameters given in the previous section, we get $\Ga(\chi \rightarrow N_R \nu_L)\sim 4\times 10^{-32}$~GeV. It leads to the lifetime of $\chi$ to be $\tau_{\chi}=\Ga^{-1}(\chi \rightarrow N_R \nu_L)\sim 0.5$ yr. Therefore, $\chi$ is enough lived to present at dark matter detectors, if it is created.   

On the other hand, in the alternative scenario with the nonthermal leptogenesis, the candidate $\chi$ can contribute to dark matter abundance if it is lighter than $N$. Indeed, in such case the generated asymmetry of $\chi$ from inflation and right-handed neutrino decay can remain to present day, since the asymmetric decay rate may always be smaller than the expansion rate, such that $\chi$ neither reaches thermal equilibrium nor thermalizes with the ordinary particles. In this case, $N$ can be long lived if it is produced by some source, similar to the above thermal case for $\chi$.

That said, the dark matter detections and collider searches should take both candidates into account. Unfortunately, the fermion candidate with gauge quantum numbers $N\sim (1,1,0,0)$ is truly sterile and does not interact with the Higgs fields $H,H'$ (perhaps it has only gravity interaction). Since the interaction of $N$ with normal matter is stringently suppressed, it easily evades the current bounds from detection experiments and colliders. Therefore, we only discuss the scalar candidate hereafter.         

The dark matter direct detection experiments measure recoil energy deposited by dark matter when it scatters off nucleons of heavy nuclei in a large detector. At the fundamental level, this scattering is due to the interaction of dark matter with quarks and gluons confined in nucleons via a portal. Since $Z'$ and $H'$ are superheavy, the viable portal is the standard model Higgs boson, $H$. Indeed, the dark matter $\chi$ scatters off nucleons via the Higgs portal determined by the interaction, $\mathcal{L}\supset -\la_5 v H\chi^*\chi$, in which $H$ interacts with quarks confined in nucleons of the nuclei as usual. Notice that the interaction of $H$ with gluons is induced by loops to be small. Given that $\chi$ is nonrelativistic, the relevant process can be determined by the effective Lagrangian, such as \cite{Belanger:2008sj}
\be \mathcal{L}_\mathrm{eff}\supset 2\la_q m_\chi \chi^*\chi \bar{q}q,\ee which has only spin-independent and even interactions. The strength of the effective interaction is induced by the $t$-channel exchange of the Higgs $H$ to be 
\be \la_q=\fr{\la_5 m_q}{2 m_\chi m^2_H}.\ee The $\chi$-nucleon ($p/n$) scattering amplitude is summed over the quark-level contributions weighted by the corresponding nucleon form factors. Thus, it takes the form, \be \sigma_{\chi-p/n}=\fr{4m^2_r}{\pi}\la^2_{p/n},\ee where $m_r=m_\chi m_{p/n}/(m_\chi+m_{p/n})\simeq m_{p/n}$, and 
\be \fr{\la_{p/n}}{m_{p/n}}=\sum_{u,d,s}f^{p/n}_{Tq}\fr{\la_q}{m_q}+\fr{2}{27}f^{p/n}_{TG}\sum_{c,b,t}\fr{\la_q}{m_q}, \ee where $f^{p/n}_{TG}=1-\sum_{u,d,s}f^{p/n}_{Tq}$ with the $f^{p/n}_{Tq}$ values given by \cite{Ellis:2000ds} \be f^{p/n}_{Tu}=0.014\pm 0.003,\hs f^{p/n}_{Td}=0.036\pm 0.008,\hs f^{p/n}_{Ts}=0.118\pm 0.062.\ee     

Taking $m_{p/n}=1$ GeV and $m_{H}=125$ GeV, the $\chi$-nucleon cross-section is evaluated to be
\bea \sigma_{\chi-p/n} &\simeq& \left(\fr{\la_5}{0.1}\right)^2\left(\fr{1\ \mathrm{TeV}}{m_\chi}\right)^2\times 6.125\times 10^{-46}\ \mathrm{cm}^2\\
&\simeq& \left(\fr{\la_5}{0.004}\right)^2\left(\fr{100\ \mathrm{GeV}}{m_\chi}\right)^2\times 10^{-46}\ \mathrm{cm}^2.\eea Provided that $m_\chi$ is at TeV as in the nonthermal leptogenesis and $\la_5$ similar to the Higgs coupling, the model predicts $\sigma_{\chi-p/n}\sim 6\times 10^{-46}\ \mathrm{cm}^2$, in good agreement with the current search \cite{Aprile:2017iyp}. Additionally, if the dark matter mass is at the weak scale, say $m_\chi\sim 100$ GeV, comparing to the data $\sigma_{\chi-p/n}\sim 10^{-46}\ \mathrm{cm}^2$ requires $\la_5\sim 0.004$. Of course, for the case of dark matter originating from the thermal leptogenesis, we should discuss a mechanism for producing $\chi$, as mentioned in footnote 7, before it reaches the XENON detector. Below, we investigate one of such mechanisms, but such search also applies for $\chi$ to be a stable dark matter.   

\subsection{Mono-$X$ signature}

At particle colliders, the scalar dark matter $\chi$ may be directly created, recognized in the form of large missing transverse momentum (or energy). An expected signal of such dark matter would be associated with a corresponding excess of a mono-$X$ or two-$X$'s final state, which recoils against such missing energy carried by the dark matter. The state $X$ might include a jet (quark, gluon), a boson (gauge, Higgs), or a lepton (charged, neutral). For simplicity and concreteness, this work considers only mono-$X$ signature and investigates such process at the LHC. 

\begin{figure}[!h]
\bc
\includegraphics[scale=0.8]{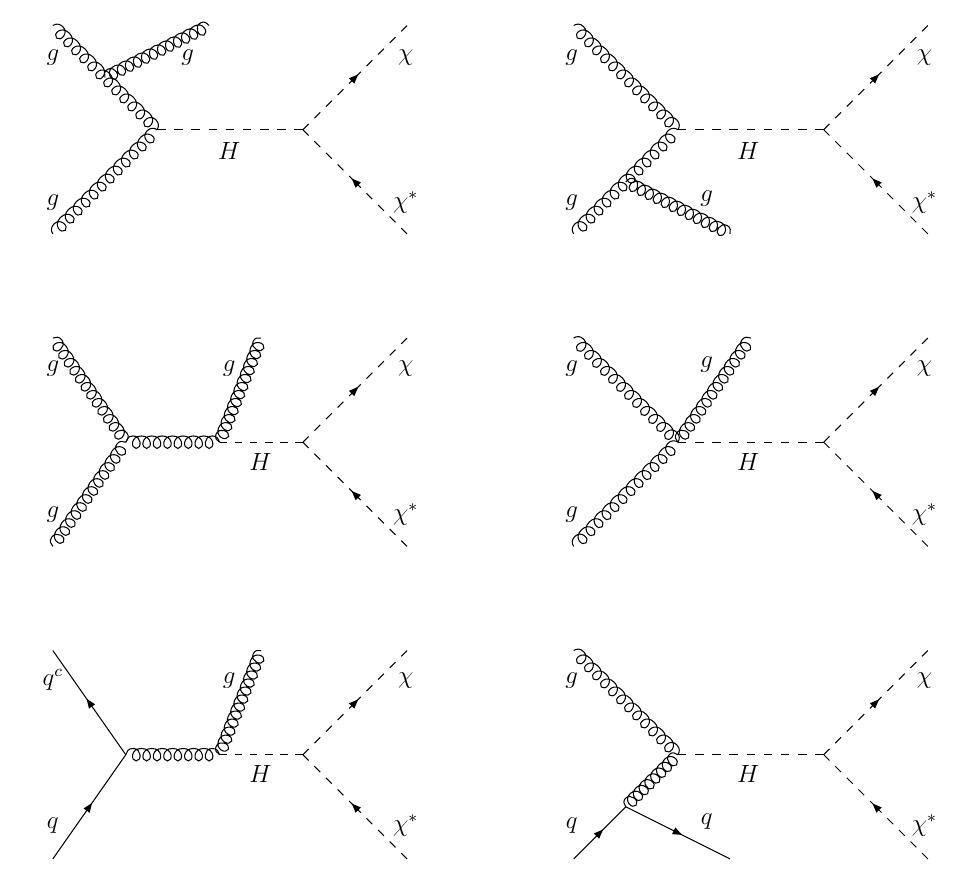}
\caption{\label{fig5} Monojet production at the LHC associated with the emission of a dark matter pair.} 
\ec
\end{figure}

\begin{figure}[!h]
\bc
\includegraphics[scale=0.8]{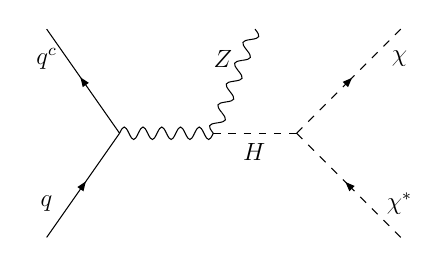}
\caption{\label{fig6} Mono-$Z$ production at the LHC associated with the emission of a dark matter pair.} 
\ec
\end{figure}

\begin{figure}[!h]
\bc
\includegraphics[scale=0.8]{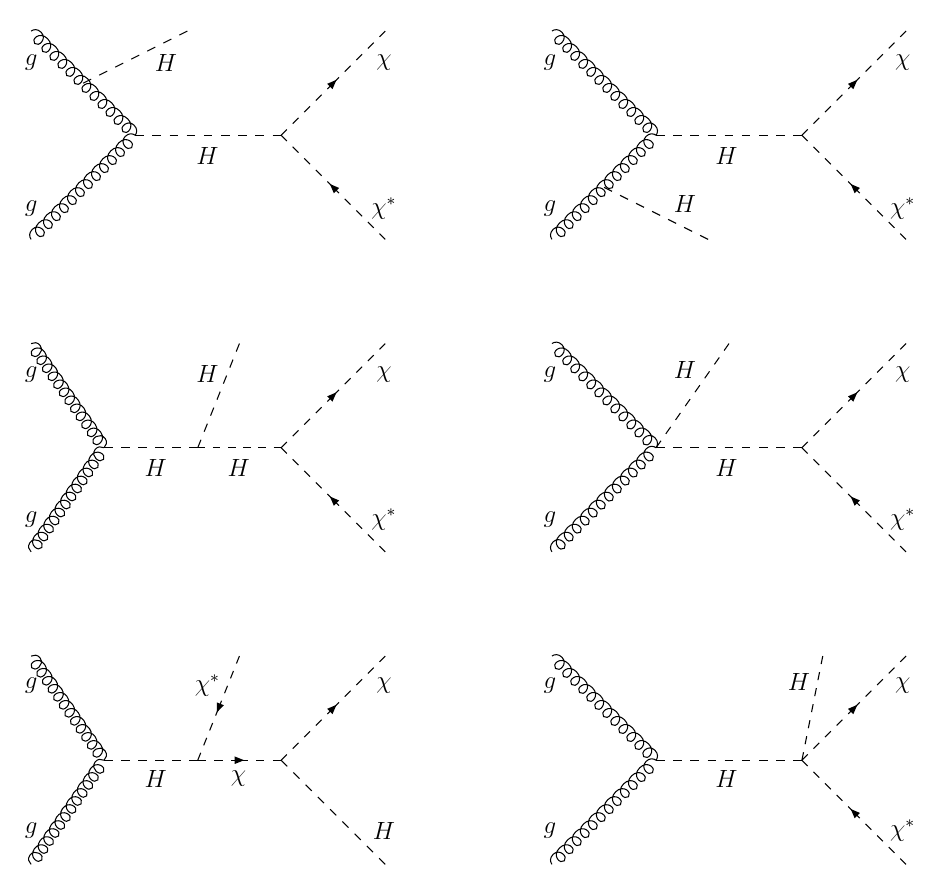}
\caption{\label{fig7} Mono-$H$ production at the LHC associated with the emission of a dark matter pair.} 
\ec
\end{figure}

In this model, the scalar dark matter and the normal matter couple through the standard model Higgs boson portal due to the interaction, $-\la_5 v H\chi^*\chi$, as mentioned, as well as those of $H$ to the standard model particles.\footnote{Because of $\La\gg v$, the mixing between $Z$ and $Z'$ due to the kinetic mixing term and the spontaneous symmetry breaking is absolutely suppressed. Hence, the $Z$ boson does not interact directly with $\chi$ and that the $Z$ portal is unavailable.} Notice that, for the latter at the LHC, the Higgs boson $H$ can couple to gluons dominantly via top loop as well as to quarks via the tree-level Yukawa interactions, as usual. Hence, through the $H$ exchange, the mono-$X$ signature contains mainly (i) a jet via the process $pp\rightarrow j\chi^*\chi$, which is given at partonic level as $gg\rightarrow g \chi^*\chi$, $qq^c\rightarrow g \chi^*\chi$, and $gq\rightarrow q \chi^*\chi$ as despited in Fig. \ref{fig5}, (ii) a $Z$ via the process $qq^c\rightarrow Z \chi^*\chi$ as presented in Fig. \ref{fig6}, and (iii) a $H$ via $gg\rightarrow H\chi^*\chi$ processes, as described in Fig. \ref{fig7}. Here the pair of dark matter produced ($\chi^*\chi$) pass in form of missing energy at the detectors, while the observable mono-$X$ signal characterizes those dark matter candidates due to the laws of conservation.      

All the processes $pp\rightarrow j \chi^*\chi$, $qq^c\rightarrow Z \chi^*\chi$, and $gg\rightarrow H\chi^*\chi$ depend on two nontrivial parameters, $\la_5$ and $m_\chi$, since the rest is known in the standard model. Additionally, all the corresponding amplitudes are proportional to $\la_5$, except for the only contribution with the left-bottom diagram in Fig. \ref{fig7}, which is proportional to $\la_5^2$. For $\la_5<1$ as investigated in the previous section, the odd contribution is negligible, meaning that all the corresponding cross-sections scale as $\la^2_5$. In other words, the dark-matter Higgs coupling $\la_5$ simply scales the production cross-sections as $\la^2_5$ and that such processes indeed depend only on one parameter, $m_\chi$, justifying a simplified model with the Higgs boson portal in order to set the LHC limits. It is clear that the dark matter candidate with a mass at TeV regime yields a signal strength more smaller than that with a mass at the weak scale. Additionally, a numerical investigation shows that the production cross-sections for mono-$Z$ and -$H$ are radically smaller than the monojet one. Hence, we display only the result for the monojet process. Generalizing the result in \cite{Belyaev:2016lok} for $m_\chi = 100$ GeV, one has \be \sigma(pp\rightarrow j\chi^*\chi)\simeq 1.8\la^2_5\ \mathrm{fb}\ \mathrm{and}\ 6.25 \la^2_5\ \mathrm{fb},\ee at the LHC for $\sqrt{s}=8$ TeV and 13 TeV, respectively. For the benchmark value of $\la_5=0.004$ corresponding to the $\chi$ mass obtained in the direct detection, the model predicts $\sigma(pp\rightarrow j\chi^*\chi)\simeq 2.9\times 10^{-5}$ fb and $10^{-4}$ fb, according to the above collision energies.                 

Last, but not least, since the new gauge $Z'$ and Higgs $H'$ bosons are superheavy, they negligibly contribute to the production cross-sections of the mono-$X$ signature. Indeed, from the gauge interactions of $Z'$ with quarks and $\chi$ as well as from the interactions of $H'$ with gluons and $\chi$ (determined through the mixing with $H$ by an angle $\sim v/\La$), integrate $Z',H'$ out. The new gauge and Higgs portals yield an effective Lagrangian \be \mathcal{L}_{\mathrm{eff}}\supset \fr{1}{12\La^2}\bar{q}\ga_\mu q \chi^*\overleftrightarrow{i\pa}^\mu \chi+\fr{\la_5\al_s}{12\pi m^2_{H'}}\fr{v^2}{\La^2}G_{n\mu\nu}G_n^{\mu\nu}\chi^*\chi.\ee Studying the mono-gluon signatures, Ref. \cite{Belyaev:2018pqr} shows the constraints on the effective couplings, namely \be \fr{1}{12\La^2} < \fr{1}{(0.3\ \mathrm{TeV})^{2}},\hs \fr{\la_5\al_s}{12\pi m^2_{H'}}\fr{v^2}{\La^2} < \fr{1}{(3\ \mathrm{TeV})^{2}}.\ee They are obviously satisfied for the choice of the parameters from the outset, $\La\gg v$, $m_{H'}=\sqrt{2\la_2}\La\gg m_H$, and $\la_5<4\pi$ limited by the perturbative condition.   

Since the fermion candidate $N_R$ interacts very weakly with the normal matter, it easily escapes from the current experimental searches, analogous to the right-handed neutrino singlet often interpreted in the literature. In other words, the dark matter $N_R$ can have an arbitrary mass above keV.    

\subsection{Alternative to asymmetric dark matter}

The XENON1T experiment has reported an excess of electronic recoil events possessing a recoil energy ranging from 1 keV to 7 keV, peaked about 2.3 keV, with a high statistical significance over $3\sigma$ \cite{Aprile:2020tmw}. Such electronic recoil signals seemingly reveal the presence of a structured dark matter. Indeed, the analysis in \cite{Kannike:2020agf} indicated that the dark matter scattering on electrons would be fast moving with velocity $v\sim 0.03$--$0.25$ for dark matter mass from 0.1 MeV to 10 GeV, which is one order larger than the typical velocity of cold dark matter $v\sim 10^{-3}$ rotating around our galaxy. This suggests a two-component structure of dark matter, one of it dominates the galactic halo in form of cold dark matter, having only gravitational interaction, whereas the other one need not to contribute to the abundance, created and boosted in annihilation of the cold dark matter, subsequently scattering on electrons in the XENON detector. Alternatively, the dark matter components have a small mass separation, such that the inelastic scattering of the higher mass component into the lower mass component with electrons may cause the excess. Indeed, the multicomponent dark matter has been phenomenologically motivated as revealing interesting consequences for galaxy structure \cite{Fan:2013yva,Fan:2013tia}, especially to accommodate the multiple gamma-ray line and boosted dark matter signals \cite{Agashe:2014yua,Kong:2014mia}, as well as dark matter self-interactions \cite{Elbert:2014bma}. This work does not go into those issues in detail, but arguing that the model provides necessary ingredients for them, in alternative to the studied scenario of asymmetric dark matter.   

Going back Table \ref{add1tb}, the model has another residual gauge symmetry, the quotient $Z_3$ group generated by \be[p^2]=[w^{3(B-L)}],\ee in addition to the matter parity. $[p^2]$ is nontrivial, if \be B-L=\pm \fr 1 3 +k=\mp \fr 1 3,\pm \fr 2 3, \pm \fr 4 3,\cdots, \ee for $k$ integer and generic field (not distinguishing fermion or boson). Thus the minimal solution is $B-L=1/3$ or $-1/3$, transforming as $\underline{1}'$ or $\underline{1}''$ under $Z_3$, respectively. We introduce two colorless fields, a vectorlike fermion $F=(1,1,0,1/3)$ and a scalar $S\sim(1,1,0,1/3)$, which match the solution. Notice that the charge $B-L=-1/3$ is characterized by the conjugated $F,S$, not necessarily introduced an extra field. The solution $S$ coincides with the bosonic solution of the matter parity. So, $S$ transforms nontrivially under both the matter parity and $Z_3$. The representations of the dark fields under these residual symmetries are given in Table \ref{faddtb9}. 
\begin{table}[h]
\bc
\begin{tabular}{c|cc}
\hline\hline
Symmetry & $W_P=(-1)^{3(B-L)+2s}$ & $[p^2]=[w^{3(B-L)}]$ \\
\hline
$N$ & $-1$ & 1\\ 
\hline
$\chi$ & $-1$ & 1 \\ 
\hline
$F$ & 1 & $w$\\ 
\hline
$S$ & $-1$ & $w$\\ 
\hline\hline
\end{tabular}
\caption[]{\label{faddtb9} Dark field representation under the complete residual gauge symmetry, the matter parity generated by $W_P$ and the quotient $Z_3$ group generated by $[p^2]$.}
\ec
\end{table} Notice that like $F,S$, the ordinary quarks transform nontrivially under $Z_3$ (see Table \ref{add1tb}). Now taking $SU(3)_C$ into account, the dark field $F,S$ are neutral, while the quark $u,d$ have 3 colors. As a result, $F,S$ cannot decay to $u,d$, if kinematically allowed/opened, due to the $SU(3)_C$ conservation. The lightest of $F,S$ is stabilized by $Z_3$ with an arbitrary mass.

The two residual symmetries define a scenario of two-component dark matter. They are 
\ben             
\item $\chi,S$: Both contribute to the dark matter relic as a co-WIMP, set by the Higgs $H,H'$ and gauge $Z'$ portals and the self-interaction $\chi^*\chi S^*S$.
\item $N,F$: Acting as co-WIMP, if $N$ is coupled to $F$ through either a singlet scalar field or a heavier $S$ dark field as added. Additionally, $F$ has $Z'$ portal, connecting to normal particles, whereas $N$ does not. 
\item $N,S$: Acting as co-WIMP, if $N$ interacts with $S$ via either a singlet scalar field or a heavier $F$ dark field imposed. In this case, $S$ has extra $H,Z',H'$ portals, while $N$ does not.  
\item $F,S$: Dark matter observables are governed by $Z'$ portal. Additionally, $S$ has extra $H,H'$ portals. 
\item $F,\chi$: Similar to the $F,S$ case.      
\een Above, the breaking of $U(1)_{B-L}$ is assumed at TeV regime. Note that $N\sim (1,1,0,0)$ only interacts with normal matter via gravity or self-interacts with other dark fields. The mechanism for producing it in the early universe is through the interaction with other dark fields, different from the leptogenesis mechanism. Additionally, the remaining dark fields $\chi,F,S$ contribute apart to the abundance (even vanished), depending on how strong they interact with the standard model particles. 
Even if one component has vanishing density, at present it can be created, boosted in the annihilation of the remaining component---the cold dark matter---through the dark matter self-interaction. Hence, the boosted dark mater is popular in this setup, worth in finding the evidence for dark matter.  

Final remark is that for the case $N,\chi$ considered in the body text and for $\La\sim$ TeV, one integrates $\nu_R$ out from the interaction $y_j \bar{N}^c_R \chi \nu_{jR}+H.c.$, leading to an effective coupling, \be \mathcal{L}_{\mathrm{eff}}\supset \fr{y^2_j}{m_{\nu_{jR}}}\bar{N}^c_R N_R\chi^*\chi+H.c.\ee The light state between $\chi$ and $N_R$ is stabilized by the matter parity, while the remainder fast decays. If $N_R$ is lighter than $\chi$, it can be produced by annihilation of $\chi$ via the above effective interaction, differing from the leptogenesis. By contrast, if $\chi$ is lighter than $N_R$, its density is set by the gauge and Higgs portals.                       

\section{\label{concl} Conclusion}

We proved that the $U(1)_{B-L}$ gauge theory provides a manifest solution for the leading questions, such as the neutrino masses and cosmological issues of inflation, dark matter and baryon asymmetry. In fact, the $B-L$ anomaly cancelation obeys the new degrees of freedom for dark matter, and the matter parity arises as a residual $B-L$ gauge symmetry, making such candidates stable. Additionally, the $B-L$ dynamics determines the neutrino mass generation seesaw mechanism, new Higgs inflation scenario when including a nonminimal interaction with gravity, as well as reheating the early universe by inflaton decays to right-handed neutrinos. The lightest right-handed neutrino of which decays CP-asymmetrically to both the present-day observed dark matter and normal matter asymmetries. The residual symmetry of the model may be larger than the matter parity, that is multiplied by a quotient $Z_3$ group, yielding interesting two-component dark matter scenarios.          

\section*{Acknowledgments}

This research is funded by Vietnam National Foundation for Science and Technology Development (NAFOSTED) under grant number 103.01-2019.353. 

\appendix

\section{\label{inflation} Inflation}

We consider the inflation scheme derived by the $B-L$ breaking scalar field, governed by the potential,
\be V(\Phi)=\fr 1 2 \mu^2_2 \Phi^2 +\fr 1 4 \la_2 \Phi^4,\ee where the inflaton field $\Phi\equiv \sqrt{2}\Re(\varphi)\simeq \La+H'$ is the real part of $\varphi$, while its imaginary part $G_{Z'}$ was absorbed to the longitudinal component of $Z'$ gauge boson by a gauge transformation, $\varphi\rightarrow \Phi/\sqrt{2}=e^{-iG_{Z'}/\La}\varphi$. 

Note that this tree-level potential is disfavored by the current data (cf. \cite{tanabashi:2018oca} for a review of inflation potential). Furthermore, $\Phi$ couples to the extra fields $Z'$, $\nu_R$, $\phi$, and $\chi$ which modify $V(\Phi)$ by a Coleman-Weinberg potential \cite{Coleman:1973jx} through quantum corrections, such as
\bea &&V_{\mathrm{CW}}(\Phi)=\fr{a}{64\pi^2}\Phi^4\left(\ln \fr{\Phi^2}{\La^2}-\fr 1 2 \right),\\ 
&& a = 9\la^2_2+\fr 1 4 \la^2_4+\fr 1 4 \la^2_6-\fr 1 2 x^4_{ii}+48 g^4_{B-L},\eea where the renormalization scale is fixed as $\langle \Phi\rangle =\La$ at which the total potential $V_{\mathrm{tot}}=V+V_{\mathrm{CW}}$ possesses a local minimum, given that $a/\la_2>-16\pi^2$, responsible for $B-L$ breaking. However, the total potential does not naturally fit the data too, since (i) the large-field inflation $\Phi>\La$ simply mimics the tree-level one, where \be V_{\mathrm{tot}}(\Phi) \simeq \fr 1 4 \left(\la_2+\fr{a}{16\pi^2}\ln\fr{\Phi^2}{\La^2} \right)\Phi^4\ee is almost insensitive to $\Phi/\La$, while (ii) the small-field inflation $\Phi<\La$ predicts a too large number of $e$-folds in contradiction to the standard cosmological evolution, as shown in \cite{Dong:2018aak}. 

In what follows, the Coleman-Weinberg contribution is skipped, i.e. $V_{\mathrm{tot}}= V$. Fortunately, when $\Phi$ rolls to the potential minimum from large value, $\Phi>\La$, the potential is approximated as 
\be V(\Phi)\simeq \fr 1 4 \la_2 \Phi^4,\ee which conserves a scale (or conformal) symmetry.\footnote{Conversely this scale symmetry suppresses the quadratic term $\fr 1 2 \mu^2_2\Phi^2$.} Including gravitational effect, the theory contains a nonminimal coupling of inflaton to gravity, \be \mathcal{L}\supset \fr 1 2 (m^2_P+\xi \Phi^2)R,\ee where $m_P=2.4\times 10^{18}$ GeV is the reduced Planck mass, $R$ is the Ricci scalar, and $\xi$ satisfies $1\ll \xi\ll (m_P/\La)^2$ in order to maintain a chaotic inflation and consistent Higgs physics from induced gravity \cite{Bezrukov:2007ep}. Conformally transforming the Lagrangian to the canonical form in the Einstein frame, $\hat{g}_{\mu\nu}=\Om^2 g_{\mu\nu}$ with $\Om^2=1+\xi\Phi^2/m^2_P$, the effective potential takes the form,
\be U(\hat{\Phi})=\fr{V}{\Om^4}\simeq \fr{\la_2 m^4_P}{4\xi^2}\left(1-e^{-\sqrt{\fr{2}{3}}\fr{\hat{\Phi}}{m_P}}\right)^2,\label{adtvl11}\ee where $\hat{\Phi}=\sqrt{3/2}m_P\ln \Om^2$ is canonically normalized inflaton field \cite{GarciaBellido:2008ab}. 

The potential $U(\hat{\Phi})$ is flat with large field values, $\hat{\Phi}\gg m_P$, yielding appropriate inflation observables. Indeed, the slow-roll parameters are directly computed from (\ref{adtvl11}) as \bea \ep &=& \fr 1 2 m^2_P\left(\fr{U'(\hat{\Phi})}{U(\hat{\Phi})}\right)^2\simeq \fr{4m^4_P}{3\xi^2\Phi^4},\\ 
\eta &=& m^2_P\fr{U''(\hat{\Phi})}{U(\hat{\Phi})}\simeq \fr{4m^4_P}{3\xi^2\Phi^4} -\fr{4m^2_P}{3\xi \Phi^2},\\ \zeta^2 &=& m^4_P\fr{U'(\hat{\Phi})U'''(\hat{\Phi})}{U^2(\hat{\Phi})}\simeq -\fr{16m^6_P}{3\xi^3\Phi^6}+\fr{16 m^4_P}{9\xi^2\Phi^4},\eea while the curvature perturbation and the number of $e$-folds are given by \bea && \Delta^2_{\mathcal{R}}=\left.\fr{U}{24\pi^2 m^4_P\ep}\right|_{k_0}\simeq \fr{\la_2\Phi^4_0}{128\pi^2m^4_P}\left(1+\fr{m^2_P}{\xi \Phi^2_0}\right)^{-2},\label{ad12a}\\ 
&& N=\fr{1}{\sqrt{2}m_P}\int^{\hat{\Phi}_0}_{\hat{\Phi}_e}\fr{d\hat{\Phi}}{\sqrt{\ep}}\simeq\fr{3\xi}{4m^2_P}(\Phi^2_0-\Phi^2_e)+\fr 3 4 \ln\fr{1+\xi\Phi^2_e/m^2_P}{1+\xi\Phi^2_0/m^2_P},\label{ad11a}\eea respectively. Here $\Delta^2_{\mathcal{R}}=2.215\times 10^{-9}$ is determined at the pivot scale $k_0=0.05\ \mathrm{Mpc}^{-1}$, while $\Phi_e$ ($\hat{\Phi}_e$) and $\Phi_0$ ($\hat{\Phi}_0$) define the $\Phi$ ($\hat{\Phi}$) value at inflation end according to $\ep=1$, i.e. $\xi\Phi^2_e=(2/\sqrt{3})m^2_P$, and at horizon exit according to $k_0$, respectively \cite{tanabashi:2018oca}. 

The $e$-folding number logarithmically depends on the inflation scale and the reheating temperature, hence possessing a value in the range $N= 50$--$60$ in order to explain the horizon problem in the standard cosmological evolution \cite{Liddle:2003as}. It follows that $\xi\Phi^2_0/m^2_P=71.33$--$84.84$ from (\ref{ad11a}) and $\xi/\sqrt{\la_2}=(4.23$--$5.04)\times 10^{4}$ from (\ref{ad12a}) according to the $N$ range, respectively. Corresponding to this $N$ range, the inflation predictions are given at the horizon exit, such that the spectral index $n_s\simeq 1-6\ep+2\eta=0.961$--$0.967$, the tensor-to-scalar ratio $r\simeq 16\ep =0.00419$--$0.00296$, and the running index $\al\simeq 16\ep\eta-24\ep^2-2\zeta^2=-(7.48$--$5.23)\times 10^{-4}$, respectively. All such predictions are in good agreement with the Planck and BICEP2/Keck Array experiments \cite{Ade:2015lrj,Array:2015xqh,Ade:2015tva}. 

Notice that the Coleman-Weinberg contributions would modify the inflation potential $U(\hat{\Phi})$. However, such corrections may be neglected, if we require $g_{B-L},x,(\la_{4,6})^{1/2}<(\la_2)^{1/4}\sim 0.1$, such that the induced beta function is small, i.e. $a \lesssim \la_2$. The last number, $0.1$, is roundly estimated from the above constraints for $\xi,\la_2$ as well as using $\La\sim 10^{15}$ GeV. 

After inflation, $\hat{\Phi}$ oscillates near the potential minimum which violates the scale symmetry. In this case, the soft $\mu_2$ term is turned on. Expanding the potential and including the soft $\mu_2$ term, we obtain \be U(\hat{\Phi})=\fr 1 2 m^2_{\hat{\Phi}}(\hat{\Phi}-\hat{\La})^2+\mathcal{O}(\hat{\Phi}^3),\ee  where we define $\hat{\La}=\xi \sqrt{3/2}\La^2/m_P$ and the inflaton mass is $m_{\hat{\Phi}}=\sqrt{\la_2/3}m_P/\xi\simeq 2.77\times 10^{13}\ \mathrm{GeV}$, taking $\xi/\sqrt{\la_2}=5\times 10^{4}$ into account. Since the higher order correction $\mathcal{O}(\hat{\Phi}^3)$ quickly vanishes after the end of inflation, $\hat{\Phi}<\hat{\Phi}_e=\sqrt{3/2}\ln(1+2/\sqrt{3})m_P\simeq 0.94m_P$, the corresponding Klein-Gordon equation for $\hat{\Phi}$ field gives an approximate solution, \be \hat{\Phi}\simeq (m_P/m_{\hat{\Phi}}t)\sin(m_{\hat{\Phi}}t)+\hat{\La},\ee as usual. It is stressed that the inflaton field rolling within the range $\hat{\Phi}\in (\hat{\Phi}_e, \hat{\La})$ would undergo a numerous oscillations after the inflation end to reach the minimum because of $\hat{\Phi}=\hat{\La}=\sqrt{3/2} \xi (\La/m_P)^2 m_P \ll \hat{\Phi}_e$, due to the constraint $\xi \ll (m_P/\La)^2$ from the outset.  

According to the leptogenesis in the text, one takes the $x$ coupling to be flavor diagonal, satisfying $x_{11}< x_{22,33}\sim g_{B-L}\lesssim 0.1$. Correspondingly, it leads to $m_{\nu_{1R}}< m_{\nu_{2,3R}}\sim m_{Z'}$, where \be m_{Z'}=2g_{B-L}(\sqrt{2/3}m_P \hat{\Phi}/\xi)^{1/2},\hs m_{\nu_{iR}}=-x_{ii}(m_P \hat{\Phi}/\sqrt{6}\xi)^{1/2}\ee are given in the Einstein frame. The concerning perturbative decay of inflaton, $\hat{\Phi}\rightarrow \nu_{1R}\nu_{1R}$, is allowed if $m_{\hat{\Phi}}>2m_{\nu_{1R}}$, implying \be \hat{\Phi}<\fr{1}{2\sqrt{2}}\fr{\sqrt{\la_2}m_{\hat{\Phi}}}{x^2_{11}}.\ee Comparing the r.h.s with $\hat{\Phi}_e$, the inflaton decays immediately or does so after several oscillations, if \be x_{11} \sim \fr{1}{\sqrt{6}}\ln^{-1/2}\left(1+\fr{2}{\sqrt{3}}\right)\la^{1/4}_2\left(\fr{\sqrt{\la_2}}{\xi}\right)^{1/2} \sim 10^{-4}.\ee Further, we require \be \hat{\Phi}>\mathrm{Max}\left\{\fr{1}{2\sqrt{2}}\fr{\sqrt{\la_2}m_{\hat{\Phi}}}{x^2_{22,33}}, \fr{1}{16\sqrt{2}}\fr{\sqrt{\la_2}m_{\hat{\Phi}}}{x^2_{B-L}}\right\},\ee such that the inflaton cannot decay to $\nu_{2,3R}$ and $Z'$. Comparing the r.h.s with $\hat{\La}$, one obtains \be x_{22,33}\sim g_{B-L}\gtrsim 10^{-2}.\ee With the choice of parameters, the decay channel $\hat{\Phi}\rightarrow \nu_{1R}\nu_{1R}$ is viable, yielding a width $\Ga=x^2_{11} m_{\hat{\Phi}}/32\pi$. This sets the reheating temperature to be \be T_R=\left(\fr{90}{\pi^2 g_*}\right)^{1/4}\sqrt{m_P \Ga}\simeq 4.4 \left(\fr{x_{11}}{10^{-4}}\right)10^{10}\ \mathrm{GeV},\ee where $g_*=106.75$ is the effective number of degrees of freedom. Since $x_{11}\sim 10^{-4}$, the reheating temperature is predicted to be $T_R\sim 4.4\times 10^{10}$ GeV.    

Compare the inflaton decay rate to the Hubble rate, i.e. $\Ga \simeq H$, where \be H=\fr{1}{\sqrt{3}m_P}\sqrt{\rho_{\hat{\Phi}}}\simeq 0.13 m_{\hat{\Phi}}j^{-1}\ee inversely depends on the number of inflaton semioscillations $j= m_{\hat{\Phi}}t/\pi$ after inflation. The inflaton undergoes $2j\simeq 26.14/x^2_{11}\sim 2.6\times 10^{9}$ oscillations, in order for their products to thermalize. This implies a long stage of preheating, a common issue raised in theories of perturbative inflaton decay. However, in this period of preheating, the nonperturbative decay $\hat{\Phi}\rightarrow Z'Z'$ may happen through broad and narrow parametric resonances \cite{Kofman:1994rk} characterized by gauge interaction $2g^2_{B-L}Z'^2\hat{\Phi}^2$. Here the effect of nonperturbative parametric resonance does not happen for fermionic products, say $\hat{\Phi}\rightarrow \nu_{2R}\nu_{2R}, \nu_{3R}\nu_{3R}$, due to the Pauli exclusion principle. The preheating of this model is worth exploring, a task to be investigated elsewhere.   

Last, but not least, the quantum gravity contribution to the inflation potential can be effectively expanded in terms of $U''(\hat{\Phi})/m^2_P$ and $U(\hat{\Phi})/m^4_P$, as shown in \cite{Linde:2005ht}. Indeed, the present model yields $U''(\hat{\Phi})/m^2_P=m^2_{\hat{\Phi}}/m^2_P\sim \la_2/\xi^2\sim 10^{-9}$ and $U(\hat{\Phi})/m^4_P\sim \la_2/\xi^2\sim 10^{-9}$, which are strongly suppressed. This justifies the effective theory of large field inflation under consideration, in agreement to \cite{Bezrukov:2007ep}.         

\bibliographystyle{JHEP}

\bibliography{combine}

\end{document}